# Strain-enhanced tunneling magnetoresistance in MgO magnetic tunnel junctions

Li Ming Loong[1], Xuepeng Qiu[1], Zhi Peng Neo[2], Praveen Deorani[1], Yang Wu[1], Charanjit S. Bhatia[1], Mark Saeys[2] & Hyunsoo Yang[1]

[1]*Department of Electrical and Computer Engineering, National University of Singapore, 117576 Singapore*

[2]*Department of Chemical and Biomolecular Engineering, National University of Singapore, 117576 Singapore*

Correspondence and requests for materials should be addressed to H.Y. (eleyang@nus.edu.sg)

**While the effects of lattice mismatch-induced strain, mechanical strain, as well as the intrinsic strain of thin films are sometimes detrimental, resulting in mechanical deformation and failure, strain can also be usefully harnessed for applications such as data storage, transistors, solar cells, and strain gauges, among other things. Here, we demonstrate that quantum transport across magnetic tunnel junctions (MTJs) can be significantly affected by the introduction of controllable mechanical strain, achieving an enhancement factor of ~2 in the experimental tunneling magnetoresistance (TMR) ratio. We further correlate this strain-enhanced TMR with coherent spin tunneling through the MgO barrier. Moreover, the strain-enhanced TMR is analyzed using non-equilibrium Green's function (NEGF) quantum transport calculations. Our results help elucidate the TMR mechanism at the atomic level and can provide a new way to enhance, as well as tune, the quantum properties in nanoscale materials and devices.**

It remains challenging to achieve high TMR values, a crucial figure of merit for MTJs, due to the ferromagnet-MgO lattice mismatch and interfacial defects, which are associated with undesirable strain,[1,2] even though TMR values reaching hundreds of percent have been achieved as a result of coherent tunneling through the MgO (001) crystalline barrier.[1,3-5] In other areas of research, strain engineering has improved the electronic properties of transistor materials by tuning the carriers' effective mass or mobility,[6-8] and was recently proposed as a method to increase the efficiency of solar cells.[9] Furthermore, strain has been studied in various other materials and devices, including multiferroic systems,[10] graphene[11] and photonics.[12] Previous experimental studies have investigated the effects of lattice mismatch on the TMR, and enhanced device performance could be achieved by optimizing film growth conditions to reduce the interface defect concentration and improve the interface lattice epitaxy.[4,13-17] Recently, theoretical work on other systems, such as a $Co_2CrAl/NaNbO_3/Co_2CrAl$ MTJ structure, has predicted the effects of strain on TMR.[18] Moreover, as the device dimensions in integrated circuits (ICs) continue to be scaled down, resulting in higher leakage power and energy density,[19] non-volatile MTJ-based memories using straintronics-based switching are emerging as a promising technology for future ICs.[20] Hence, as understanding how and why strain affects TMR could enable better device engineering and improve the performance of MTJs, we perform a detailed investigation of the effects of strain on MTJs.

In this work, we study the strain-enhanced TMR in MgO magnetic tunnel junctions and it is found that the conductance for the antiparallel configuration, in which two ferromagnets are aligned in opposite directions, is significantly more sensitive to strain. Measurements for the non-annealed MTJ devices and metallic GMR spin valve show that the observed strain-enhanced TMR is correlated with the coherent quantum tunneling process through the MgO barrier. Our findings can contribute towards the general understanding of the TMR mechanism at the atomic level, as we analyze the effect of different perturbations imposed on the device, and propose an explanation based on the strong dependence of the transport pathways on the relative magnetization orientation of the electrodes. Moreover, our results can provide a new way to enhance, as well as tune, the quantum properties in nanoscale materials and devices.

## Results

**Strain-enhanced TMR.** Figure 1(a) shows the $Co_{40}Fe_{40}B_{20}$/MgO/$Co_{40}Fe_{40}B_{20}$ MTJ film structure, which was deposited at room temperature by magnetron sputtering. All the film layers were deposited using dc sputtering, except the MgO tunnel barrier and $SiO_2$ encapsulation, which were deposited using rf sputtering. The sputtering pressures for the different layers were in the range of 1 – 3 mTorr. The MTJ film stack was patterned by photolithography and Ar ion milling to form MTJ junctions with sizes ranging from 36 – 10600 $\mu m^2$. Where necessary, the MTJ devices were post-annealed in a magnetic field of 0.055 T under ultra-high vacuum conditions, either at 250 ºC for 1 hour, or at 350 ºC for 10 minutes followed by 400 ºC for 30 minutes. A CIP-GMR film stack of Si/Ta (5 nm)/Ru (5 nm)/$Co_{70}Fe_{30}$ (4 nm)/Cu (2.6 nm)/ $Co_{70}Fe_{30}$ (4 nm)/$Ir_{22}Mn_{78}$ (10 nm) was deposited at room temperature by dc magnetron sputtering in an ultra-high vacuum chamber. A magnetic field of 0.055 T was applied during the deposition to exchange-bias the pinned layer.

TMR as well as I-V measurements were performed on the MTJs at room temperature using the four probe measurement technique, without and with the application of strain. The MTJs were switched using field induced magnetic switching. Strain was applied by clamping the edges of the sample and subsequently turning a screw, which pushed the center of the sample upwards, as illustrated in Fig. 1(b) and 1(c). The sample was approximately 1 cm × 0.5 cm with a thickness of 500 μm. The top surface of the sample contained 24 equally-spaced devices. The screw and clamps were made of a non-magnetic aluminum alloy, in order to avoid any magnetic influence of the setup on the measurements.

Clamping the edges of the sample enhanced the TMR (Fig. 1(d)) by a mean factor of 1.32, on average. By turning the screw, the TMR was increased further, typically by a factor of 1.27 times relative to the clamped structure, on average, yielding a mean overall enhancement of approximately 1.68 times. Figure 2(a) shows the effect of strain on the TMR of a specific 66 μm$^2$ junction. When the strain was gradually decreased and increased again, the TMR trend was reproducible, as shown in Fig. 2(b). The reversibility and reproducibility suggest that the applied strain was elastic, and no permanent deformation occurred. The highest enhancement factor of 2.25 times was observed for this device. It should be noted that in Fig. 2(b), there is some variation in the TMR values at each level of strain. This is attributed to slight but unavoidable shifting of the sample position in the clamps when strain is applied and removed. This results in a margin of error for the applied strain. However, if the devices are strained excessively, or continue to be strained repeatedly (mechanical fatigue), it is anticipated that severe plastic (permanent) deformation will eventually occur in the films, such as dislocation motion in the tunnel barrier, which could also affect the TMR. Figure 2(c) shows the voltage-dependence of the junction resistance for the same device, as the strain increases. Consistent with the typical behavior of MgO based MTJs, the resistance of the parallel state ($R_P$) was virtually independent of bias voltage, while the resistance of the antiparallel state ($R_{AP}$) decreased as bias voltage increased.[13,21,22] As the strain was increased, $R_{AP}$ increased. This enhancement of TMR saturated as the screw rotation approached 12º. Figure 2(d) shows the TMR trend for a device with a junction area of 1600 μm$^2$. The sample was strained until fracture occurred at approximately 23º of screw rotation, corresponding to biaxial strain of approximately 0.2% along the x-direction and 0.4% along the y-direction. As the strain increased, the TMR increased and eventually saturated, then dropped abruptly after fracture. The sharp reduction in TMR can be attributed to mechanical damage sustained by the device, due to the fracture of the substrate or film.

The TMR versus magnetic field curves of the devices also became increasingly square with strain (Fig. 2(a)), due to the magnetostriction effect.[23,24] The coercivity of the hard ferromagnetic layer increases with strain, while the coercivity of the soft layer remains virtually unchanged (Fig. 3(a)). The coercivities were estimated from the maxima and minima of the first derivatives of the TMR loops. This effect resulted in wider TMR loops with sharper switching (Figs. 2(a) and 4(b)). The magnetic hysteresis loop obtained using vibrating sample magnetometry (VSM) indicates that the bottom $Co_{40}Fe_{40}B_{20}$ layer is magnetically hard, while the top layer is soft (Fig. 3(b)). It has been reported that the bottom CoFeB layer undergoes more Ta diffusion,[4] therefore, the bottom CoFeB layer contains more domain wall pinning sites than the top layer, resulting in a higher coercivity and a higher magnetostriction coefficient.

**Correlation with coherent tunneling.** In order to verify that the enhanced TMR is mainly due to the effect of strain on the quantum transport properties through the crystalline MgO tunnel barrier, the experiment was repeated using non-annealed MTJ devices, which were otherwise identical to the annealed MTJ devices. Figure 4(a) shows that the bias voltage dependence of the resistances remained virtually unchanged with the application of strain, in contrast to Fig. 2(c). Moreover, Fig. 4(b) shows that the TMR remained almost constant at 15% even under strain, though the loop squareness was enhanced by strain due to the magnetostriction effect, similar to Fig. 2(a). As annealing improves the crystallinity of the MgO barrier and crystallizes the CoFeB ferromagnetic layers, thereby facilitating coherent tunneling,[4,25] the negligible dependence of the TMR on the applied strain for the non-annealed devices suggests that a lack of coherent tunneling is correlated with a TMR which is independent of strain. It implies that the strain-enhanced TMR is caused by a change in the quantum transport properties. This was verified by repeating the experiment on the same devices after annealing at 250 °C, the threshold temperature for crystallization of CoFeB,[26] for 1 hour. After annealing, the devices again displayed strain-enhanced TMR (Fig. 5). High resolution TEM images of the MTJ stack before and after annealing verify the onset of crystallization in the CoFeB layers after heating the device at 250 ºC (Fig. 6).

The experiment was furthermore repeated using a $Co_{70}Fe_{30}/Cu/Co_{70}Fe_{30}$ current-in-plane GMR (CIP-GMR) sample. As the GMR spacer was metallic, coherent tunneling through the MgO barrier was eliminated. The GMR remained virtually constant at 3% under strain, though the squareness of the GMR loop increased due to magnetostriction (Fig. S1).

**Computational methods.** To further elucidate the effect of biaxial *xy* strain on the TMR, NEGF quantum transport calculations with an extended Huckel Hamiltonian, as implemented in Green,[27-29] were performed. The tunneling junction was modeled using a 6-layer MgO(001) barrier contacted by semi-infinite Fe(100) leads (Fig. 7(a)). The electronic structure of the tunneling junction was described by an extended Hückel molecular orbital (EHMO) Hamiltonian.[30] Standard EHMO parameters were used for Fe, Mg, and O.[31] With these parameters, the calculated MgO bulk band gap of 7.8 eV agrees with the experimental value of 7.77 eV,[32] and the calculated Fe magnetic moment of 2.0 $\mu_B$ agrees with values from hybrid DFT-HSE03[33-35] calculations. Since DFT-HSE03 calculations for a 4-layer MgO film on a 6-layer Fe(100) slab shows that the top of the MgO valence band edge lies 4.0 eV below the Fermi level of the system, a similar offset was used in the transport calculations.

The calculations used the experimental Fe lattice constant, 2.86 Å, for the unstrained calculations, and a Fe(100)-MgO(001) distance of 2.16 Å.[36] Strain was modeled by stretching the x and y lattice between 0 and 1%, and the resulting compression in the z-direction was obtained from the Poisson ratios of 0.19 for MgO and 0.37 for Fe.[37] The same EHMO parameters were used for the strained structures since EHMO parameters are transferable and account for changes in the orbital overlap when strain is applied. k-resolved T(E) spectra were computed for a high resolution 100×100 grid, using a small value ($10^{-6}$ meV) for the imaginary part of energy.

Figure 7(b) shows the change in the conductance calculated for the parallel and antiparallel configurations. In agreement with the experimental results, the conductance for the parallel configuration is robust and changes little with strain. The conductance for the antiparallel configuration is significantly more sensitive to strain, and decreases gradually by 6.3% for 1% $xy$ biaxial strain. A similar trend was observed experimentally, though the experimental change is larger (Fig. 2(c)). The decrease in the antiparallel conductance increases the calculated optimistic TMR ratio by 8% for 1% $xy$ strain. The higher sensitivity of the antiparallel configuration can be understood from the $k_{//}$-resolved transmission spectra $T(E_F)$ (Figs. 7(c) and 7(d)). In agreement with previous studies,[13,36,38,39] the parallel majority conductance is dominated by states near the Γ point, while the antiparallel conductance is dominated by states away from the Γ point. As reported previously,[13] the latter transport channels depend strongly on the shape of the $E(k)$ dispersion relations near the Fermi level and are sensitive to changes in the junction structure and in the orbital overlap due to strain.

Next, we analyze the contributions to the changes in conductance for a $xy$ biaxial strain of 0.5% (Table 1). Stretching the $xy$ lattice causes a contraction in the $z$-direction for both the Fe layers and the MgO barrier, given by the Poisson ratios. Reducing the Fe lattice in the $z$-direction significantly decreases both the parallel and the antiparallel conductance by 3.6 and 6.6%, respectively. The reduced lattice increases the overlap of the $d$-orbitals and broadens the $d$-band.[13] Analysis of a Fe/vacuum/Fe junction shows that compression moves minority states away from the Γ point and reduces the conductance. This effect is amplified by the filtering effect of the MgO barrier for the minority states. Subsequent compression of MgO in the $z$-direction decreases the barrier thickness and increases the conductance. The effect is larger for the antiparallel configuration because of the larger decay coefficient.[13,36,38,39] Finally, increasing the $xy$ lattice affects the conductance only slightly. Again, the small change results from compensation. An increase of the $xy$ lattice increases the conductance for a Fe/vacuum/Fe junction by 10% (parallel) and 3% (antiparallel) because of the change in the Fe(100) Fermi surface. However, *xy* expansion also increases the MgO bandgap and changes the position of the bands, which compensates the initial effect. Table 1 demonstrates that the small *overall* change in the parallel conductance results from cancellations between these factors. The filtering effect of the MgO barrier, however, increases the effect of the change in the minority Fe(100) interface states on the antiparallel conductance, and this effect is not fully compensated by changes in the MgO band structure and barrier thickness.

**Discussion**

While the NEGF modeling results match the experimental trend of TMR with strain, the experimentally observed increase in TMR is still larger than that predicted by the modeling. Hence, other strain-induced mechanisms may also contribute to the TMR enhancement, but are not taken into account in the modeling. A plausible mechanism could be the rotation of tilted crystal planes to more favourable orientations for coherent tunneling. In real devices, the crystal structure of the MgO tunnel barrier is often imperfect, and crystal plane tilting is sometimes observed,[3] as shown in Fig. S2(a). The MgO barrier crystal planes in the boxed region on the left are tilted ~7º clockwise relative to those in the boxed region on the right. If strain is applied in any direction that is not parallel to the axis about which the plane is tilted (for example, in Fig. S2(b), the tilt axis is perpendicular to the page), then some of the crystal planes will rotate. In this study, the biaxial in-plane tensile strain would cause the orientation of tilted crystal planes to become more parallel to the substrate surface. The strain would effectively straighten the tilted planes without affecting the orientation of planes that were already parallel to the substrate. Therefore, the crystal structure of the MTJ would be closer to the ideal lattice for coherent tunneling,[36] possibly resulting in a higher TMR ratio.

It should be noted that the applied strain also improved the AP alignment of the ferromagnetic electrodes due to the magnetostriction effect, as well as reduced the domain wall (DW) concentration, thereby reducing spin mixing,[40] and thus, possibly increasing the TMR. Hence, we have performed further simulations and experiments to investigate the contributions of these different factors to TMR enhancement (Figs. S3, S4 and S5 and Table S1). Our results show that regardless of poor AP alignment and high DW concentration, strain-enhanced TMR can still occur.

In conclusion, we have observed strain-enhanced TMR using crystallized CoFeB/MgO/CoFeB MTJ structures. The trend was analyzed using NEGF quantum transport calculations, which show that the applied strain has little effect on the parallel conductance, while reducing the antiparallel conductance. Furthermore, the effect is likely enhanced by strain-induced rotation of the MTJ crystal planes towards orientations that are more favourable for coherent tunneling.

## Methods

**Finite element analysis simulations.** The vertical displacement of the center of the sample (Fig. 1(b) and 1(c)) was estimated based on the angle of screw rotation. Finite element analysis (FEA) was then used to estimate the stress and strain distribution in the sample. LISA 7.7. was used for the FEA simulations. In the FEA analysis, only the silicon substrate was modeled, as the overall MTJ film structure is relatively thin compared to the 500 μm-thick substrate. Furthermore, the Young's modulus of thin films is known to approach that of the substrate.[41] It was assumed that the clamped edges are rigid, which was modeled by zero vertical displacement boundary conditions at the clamped regions. In the FEA simulation, the Poisson's ratio was assumed to be 0.25 and the Young's modulus of silicon along the [100] crystallographic direction (129.5 GPa)[42] was assumed to be the isotropic Young's modulus of the substrate. Moreover, the contact area between the screw and the substrate was assumed to be approximately circular, with a radius of 0.5 mm.

The FEA results suggest that the devices were subjected to biaxial tensile strain, which was highest at the central stress application point, and decreased with distance away from this point. For 12º of screw rotation, the biaxial tensile strain at the central stress application point was estimated to be 0.1% along the x-direction and 0.2% along the y-direction. The relationship between strain and screw rotation angle is approximately linear. Note that this model does not account for strain introduced to the thin films by the clamping. As the bottom electrode and SiO$_2$ encapsulation layer (Fig. 1(a)) are continuous over the entire surface, including the clamped regions, and thin films are typically less dense,[43] and thus, more compressible than their corresponding bulk materials, clamping the sample compressed the aforementioned continuous thin film layers along the *z*-direction, and this strain, which was also transferred to the encapsulated devices, resulted in a corresponding expansion along the *x*- and *y*-directions, thus enhancing the TMR.

**Acknowledgements:** Discussions with Dr. Hiroyo Kawai and Ravi Tiwari, as well as TEM assistance from Shreya Kundu are gratefully acknowledged. This research is supported by the National Research Foundation, Prime Minister's Office, Singapore under its Competitive Research Programme (CRP Award No. NRF-CRP 4-2008-06). MgO targets were supplied by Ube Material Industries, Ltd.

**Author contributions:** L.M.L., C.B., and H.Y. planned the study. L.M.L. and X.Q. prepared samples and fabricated devices. L.M.L. measured transport properties and did FEA simulations. Z.N. and M.S. carried out NEGF modeling. P.D. performed OOMMF simulations. Y.W. carried out SMOKE measurements. All authors discussed the results and wrote the manuscript. H.Y. supervised the project.

**Supplementary Information** accompanies this paper on www.nature.com/scientificreports

**Competing financial interests:** The authors declare no competing financial interests.

**Figure 1| Sample structure and experimental setup.** (a) The film structure of the MTJs. (b) Finite element analysis results for the vertical displacement of the sample for 12º of screw rotation, with an amplification factor of 53 to exaggerate the deformation for clarity. (c) Side view of the setup used to apply mechanical strain, consisting of clamps and a central vertical screw. (d) Strain-enhanced TMR due to clamping, where the measurements were repeated thrice without clamping, and then thrice with clamping, to verify the effect.

**Figure 2| The effect of strain on the TMR of annealed devices.** (a) The effect of different levels of strain on the TMR loops. (b) The reversibility and reproducibility of the TMR trend during testing. The strain was increased, decreased, and then increased again, denoted by "Forward 1", "Backward", and "Forward 2" respectively. (c) The bias voltage-dependence of the junction resistance at different levels of strain. (d) The TMR trend for another device, as the substrate was strained until fracture.

**Figure 3| The effects of strain on coercivity.** (a) The coercivity of the hard and soft magnetic layers of the device used in Fig. 2(d) as a function of screw rotation angle, for forward and backward sweeps of the magnetic field. (b) The VSM data for patterned devices, indicating that the thicker $Co_{40}Fe_{40}B_{20}$ layer (the bottom ferromagnetic layer) was the hard magnetic layer.

**Figure 4| Correlation with coherent tunneling (the effect of strain on the TMR of a non-annealed MTJ device).** (a) The bias voltage-dependence of the non-annealed junction resistance at different levels of strain. (b) The TMR loops for various strain conditions. The screw was rotated incrementally from 0º to 12º, and then restored to 0º again.

**Figure 5| Strain enhanced TMR in annealed devices.** (a) The effect of strain on the TMR loops of a device after the annealing at 250 °C. (b) The TMR as a function of strain for another device, before and after the annealing.

**Figure 6| High resolution TEM images of the MTJ stack.** (a) Without annealing, negligible crystallization of the $Co_{40}Fe_{40}B_{20}$ layers was observed. (b) After the annealing at 250 ºC for 1 hour in a magnetic field of 0.055 T, crystallization of the $Co_{40}Fe_{40}B_{20}$ layers is noticeable.

**Figure 7| Non-equilibrium Green's function simulations.** (a) Fe/6-layer MgO/Fe model used in the quantum transport simulations. The Fe(100) contacts are semi-infinite. (b) Percentage change in the parallel ($1.7\times10^{-5}$ $e^2/h$) and antiparallel ($9.1\times10^{-8}$ $e^2/h$) conductance, and optimistic TMR ratio as a function of biaxial *xy*-strain. (c) $k_{//}$-resolved transmission coefficients, $T(E_F)$, for majority states, parallel configuration and (d) antiparallel configuration for the unstrained junction.

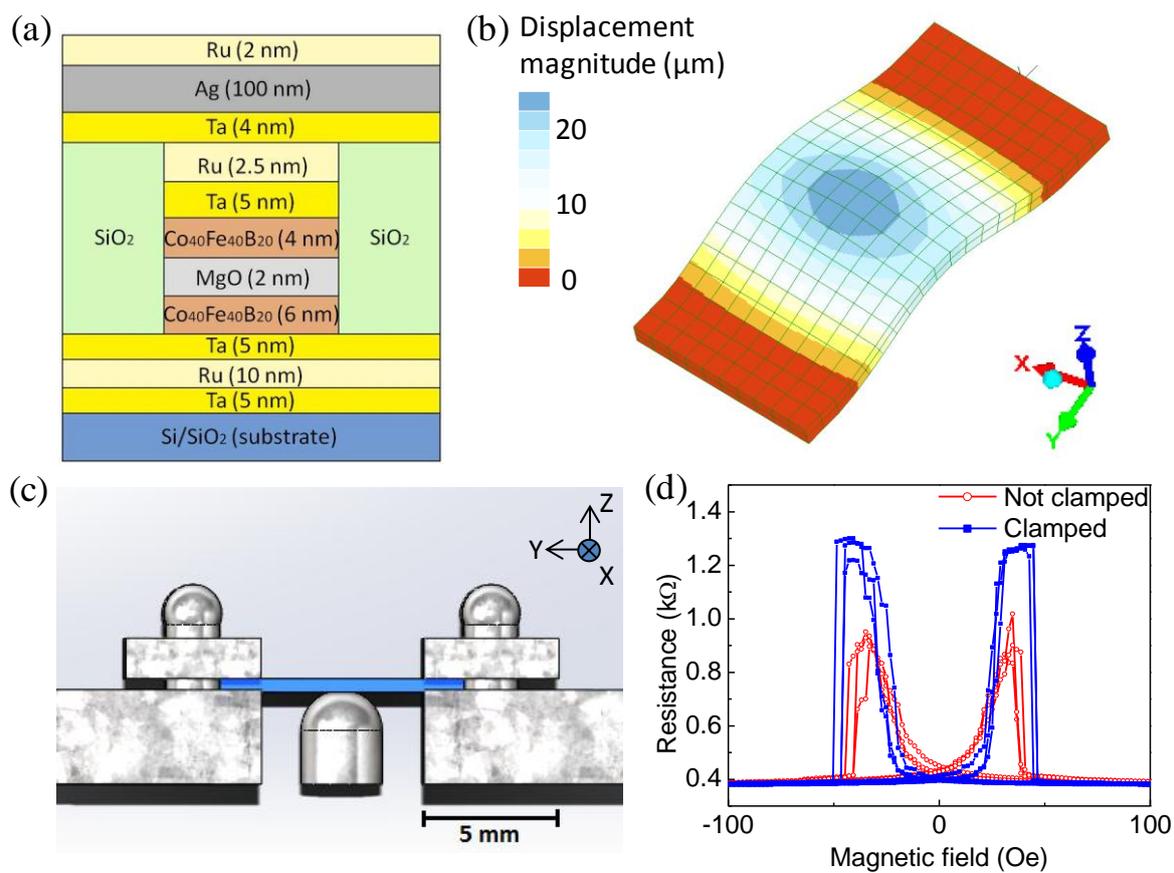

Figure 1

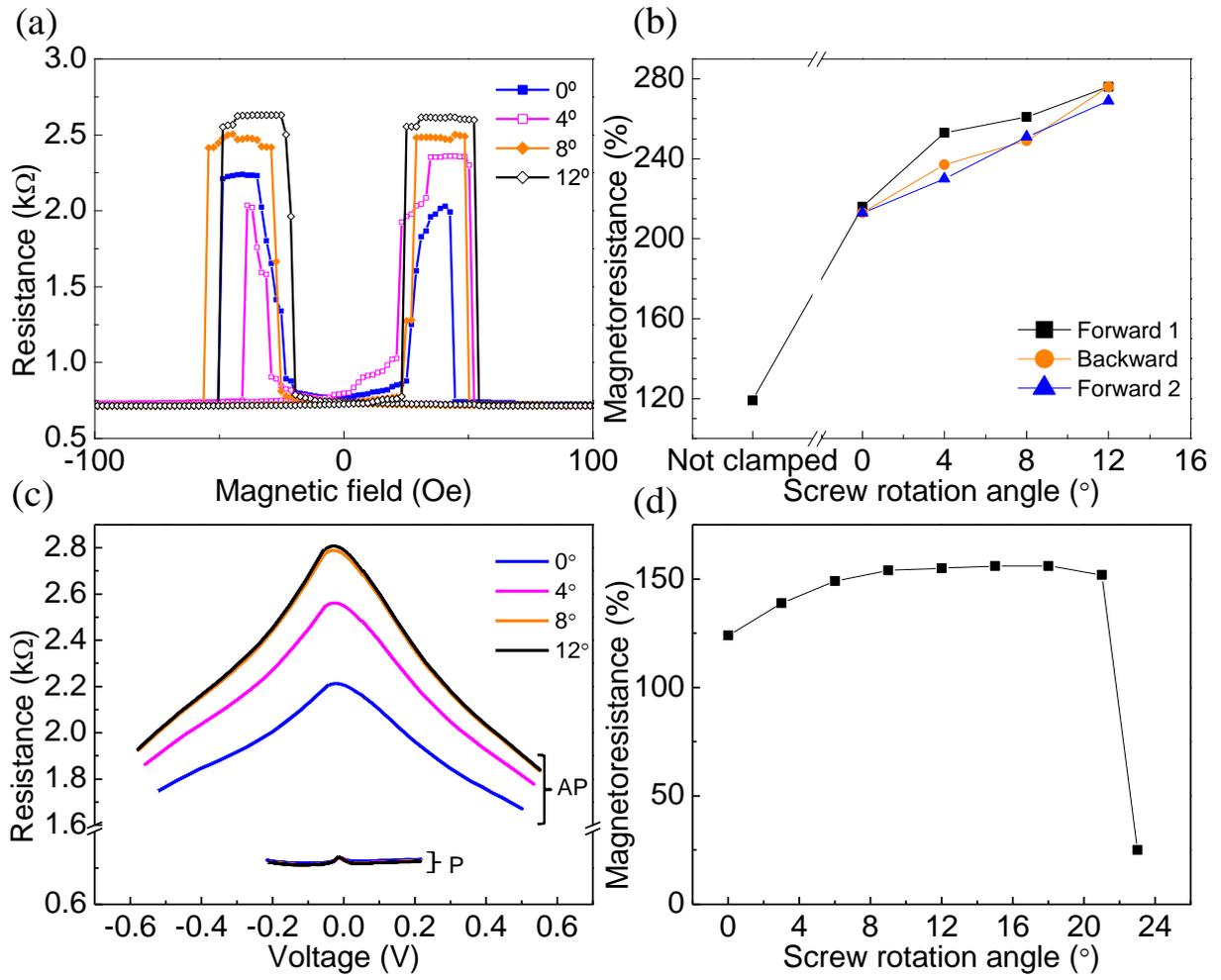

Figure 2

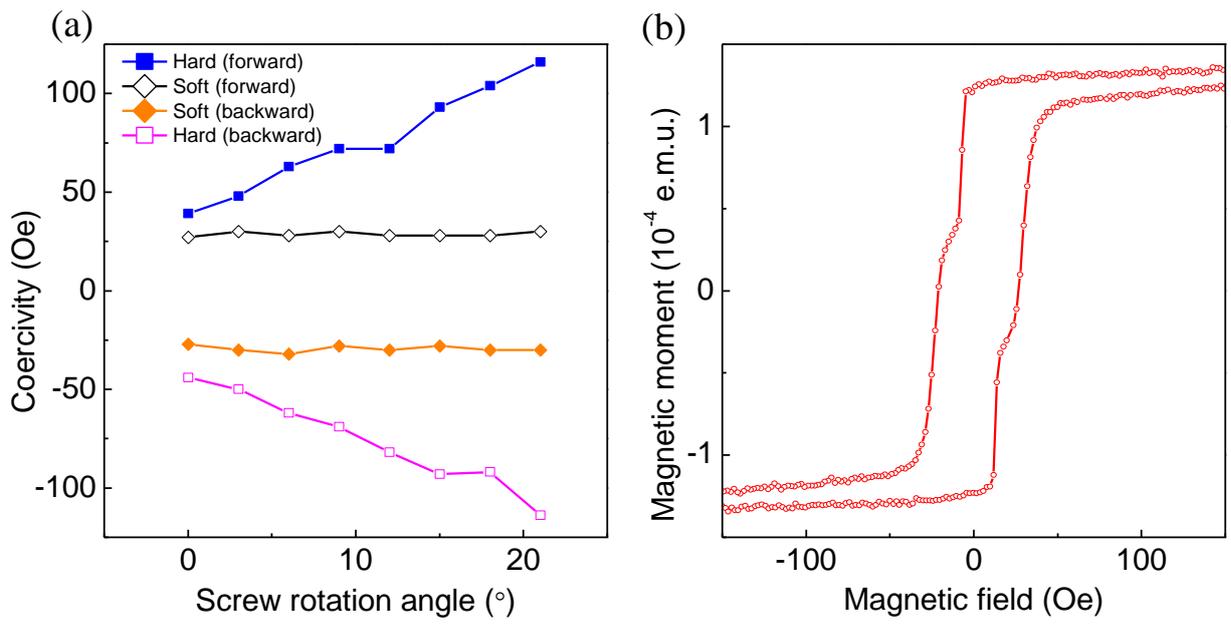

Figure 3

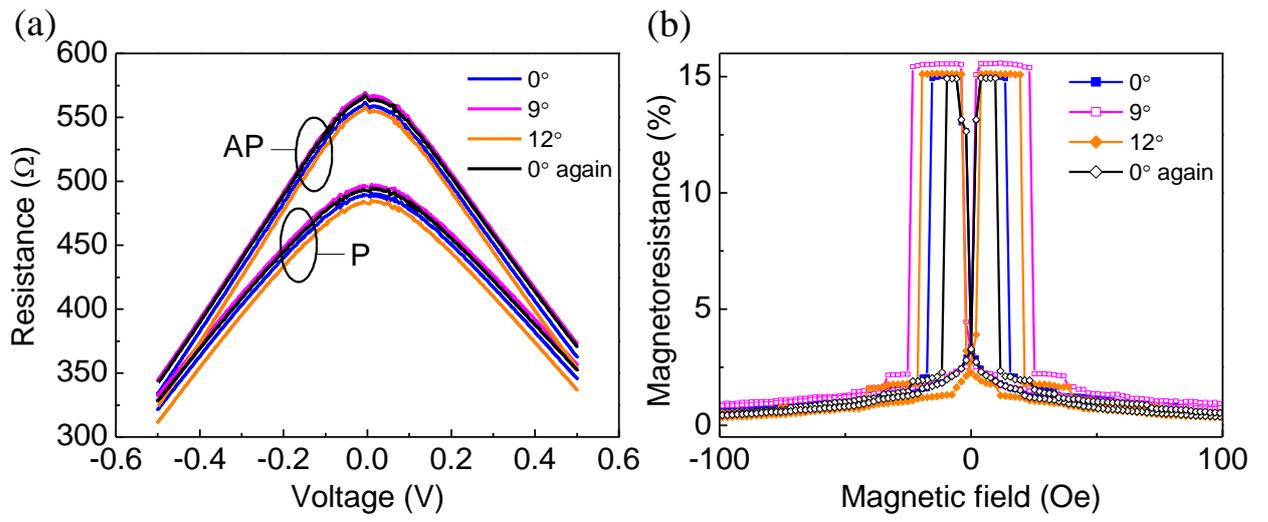

Figure 4

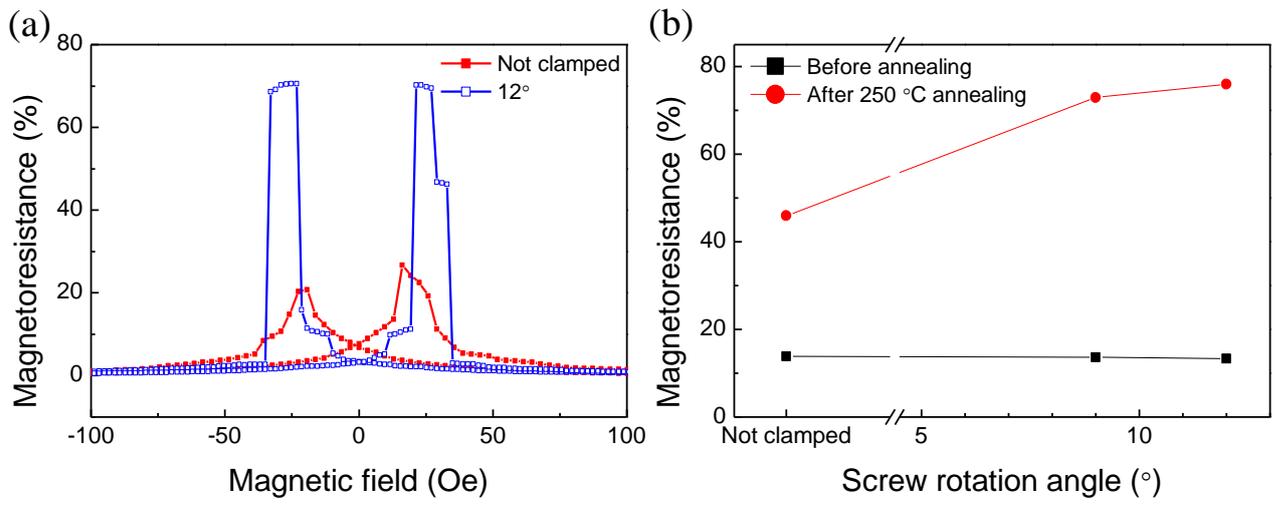

Figure 5

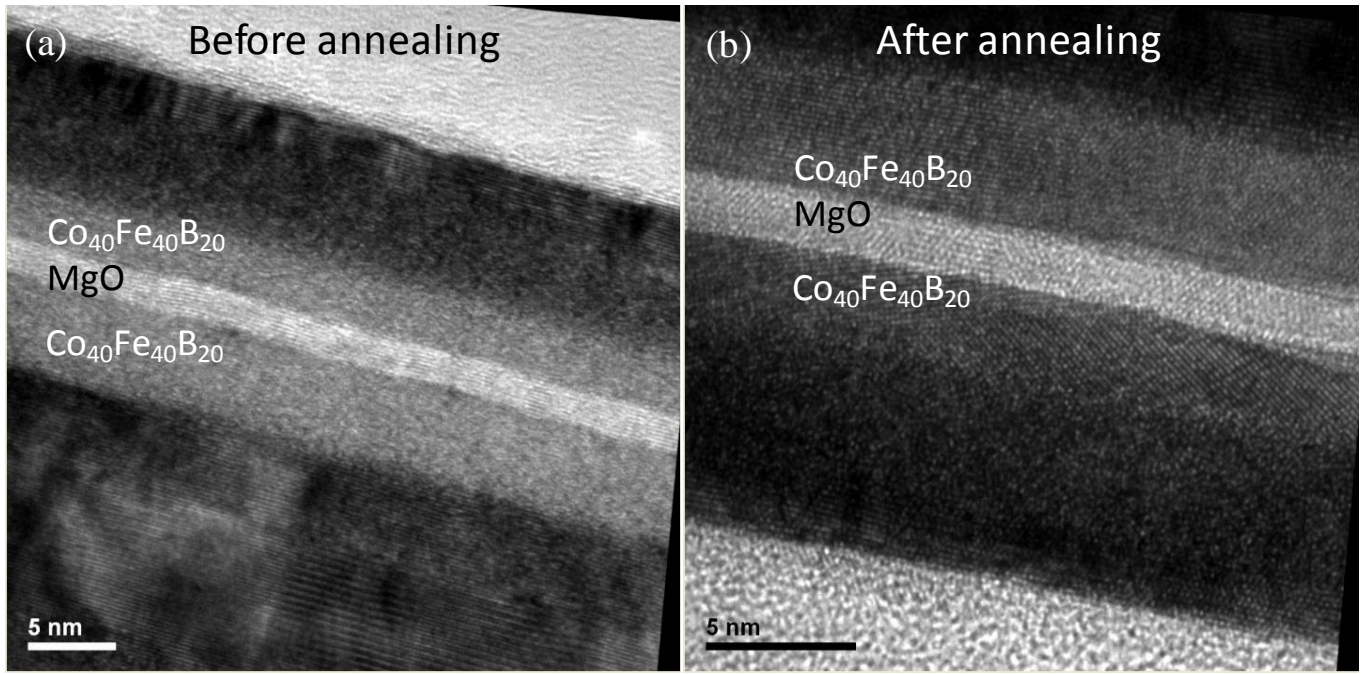

Figure 6

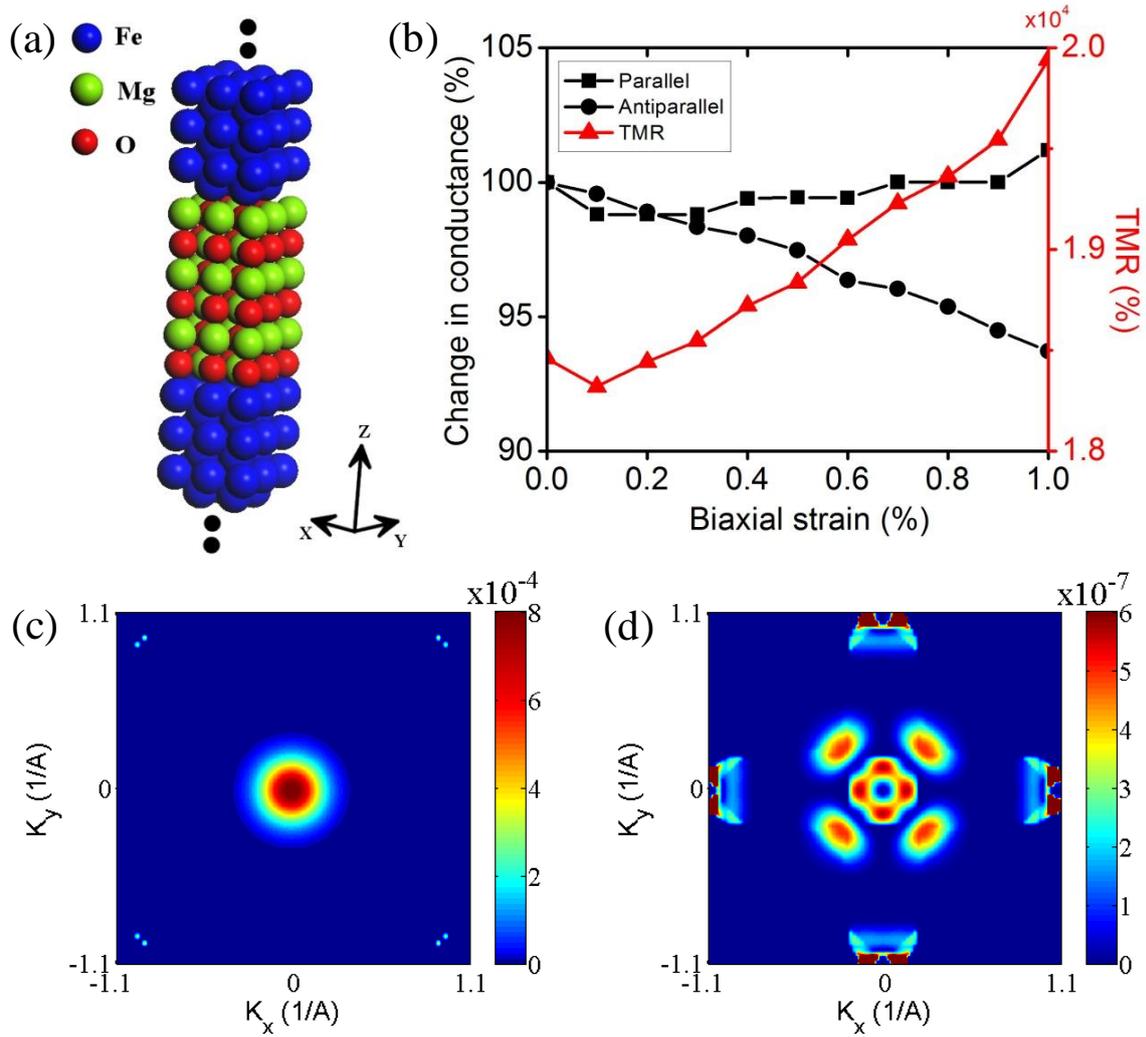

Figure 7

**Table 1** Analysis of the effect of lattice changes introduced by 0.5% *xy* biaxial strain on the parallel and antiparallel conductance for a Fe/6-layer MgO/Fe junction. The combined effect of a reduction in the *z*-lattice of the Fe contacts (i), of the MgO barrier (ii), and an expansion of the *xy* lattice of the entire junction (iii) decreases the parallel conductance by 0.7% and the antiparallel conductance by 2.5%.

|  |  | Parallel | Antiparallel |
|---|---|---|---|
| Unstrained junction conductance ($e^2/h$) | | $1.7\times10^{-5}$ | $9.1\times10^{-8}$ |
| Stepwise change in conductance | 0.58% *z*-compression of the Fe contacts | -3.6% | -6.6% |
| | 0.23% *z*-compression of the MgO barrier | +1.2% | +3.8% |
| | 0.50% *xy* expansion of the entire device | +1.8% | +0.6% |